# Energy flow lines as light paths: a didactical analysis

# Energiestromlinien als Lichtwege: Eine didaktische Analyse

Martin Erik Horn

University of Potsdam, Physics Education Research Group,
Am Neuen Palais 10, 14469 Potsdam, Germany
E-Mail: marhorn@rz.uni-potsdam.de
mail@martin-erik-horn.de

**Abstract**

Analyses of interviews with secondary school students about their conceptions of light at the University of Potsdam indicate that numerous students have a deterministic view of light. With regard to these results the model of energy flow lines, which has been discussed recently in the didactical literature, is of special interest. Following this model, light is presumed to move along energy flow lines as trajectories.

In an analysis of the model of energy flow lines four didactical dimensions (didactical content, internal structure, present-day relevance and future significance) are investigated. It can be shown that a discussion of this model in physics at school can increase the meta-conceptional knowledge of the students about the models of light. On the other hand, this can promote deterministic conceptions and the Bohm interpretation of quantum mechanics. But the question remains: Should the nature of light really be described as deterministic?

**Kurzfassung**

Die Auswertung von Schülerinterviews zu den Modellvorstellungen des Lichts, die an der Universität Potsdam durchgeführt wurde, zeigt, dass zahlreiche Schülerinnen und Schüler eine deterministische Vorstellung zum Licht besitzen. Unter Bezug auf diesen Befund ist das Modell der Energiestromlinien, das in den letzten Jahren in der Literatur diskutiert wurde, von besonderem Interesse. Im Rahmen dieses Modells werden die Energiestromlinien als Trajektorien betrachtet, denen das Licht folgt.

Diese Modellvorstellung wird hinsichtlich der vier Zieldimensionen (Gehalt, innere Struktur, Gegenwartsbedeutung und Zukunftsbedeutung aus Sicht der Lernenden) didaktisch analysiert. Es zeigt sich, dass eine fundierte Behandlung im Physikunterricht einerseits das metakonzeptuelle Verständnis für Lichtmodelle fördern kann, andererseits Vorstellungen in Sinne einer Bohmschen Deutung der Quantenmechanik induziert werden. Letztendlich stellt sich die Frage: Soll das Verhalten des Lichtes tatsächlich als deterministisch dargestellt werden?

**Contents**

**Inhalt**

## 1. Models of light

When investigating and discussing the nature of light students in school, college and university are confronted with numerous models. Among other models, light can be treated in the context of the ray model, the wave model, the phasor model, the Newtonian particle model and the quantum model.

In addition to these models, the model of energy flow lines was suggested and intensely discussed in literature (Wünscher et al. 2002) recently. Some consequences of the use of this model of light at school will subsequently be presented in this paper.

## 2. Student conceptions of light

Physics education research at the University of Potsdam has shown that using a plenitude of different models in school can lead to numerous severe complications at the process of model construction of students (Mikelskis et al. 2002). Especially in optics, the students investigated often show a tendency to construct hybrid models. In their descriptions, an appropriate und necessary distinction between different models of light is lacking (Horn et al. 2002).

In addition, students prefer descriptions of optical processes using the ray model with clearly predetermined light paths.

**Student 1**: „I always imagine photons as small particles, which are just quanta going straight through the slit.
And maybe, they slightly hit the upper edge of the slit and maybe, they are therefore slightly deflected from their straight flight path."

**Student 2**: „It will of course be especially simple if you imagine light as small particles.
(...) There the properties of particles are inherent, and they then propagate in the form of waves."

***Fig. 1:*** *Comments made by secondary school students from an advanced physics course (grade 13) on the nature of light.*

This deterministic attitude can explicitly be found in comments made by students on light, indicating a narrow association of particle paths to light particles (see figure 1). On the basis of their everyday experiences by far most students share these views.

The great attractiveness of this concept is also revealed by the high willingness of students to construct hybrid models of light. These unphysical models contain paths or directions of particle movements as essential parts of their description. The des-

## 1. Die Modelle des Lichts

Bei der Diskussion und Erörterung der Natur des Lichtes werden Schüler und Studenten mit einer Vielzahl von Lichtmodellen konfrontiert. So kann Licht unter anderem im Kontext des Strahlenmodells, des Wellenmodells, des Zeigermodells, des Newtonschen Teilchenmodells und des Quantenmodells behandelt werden.

In letzter Zeit wurde darüber hinaus das Modell der Energiestromlinien als ein didaktisch wirksames Lichtmodell vorgeschlagen und verstärkt in der Literatur diskutiert (Wünscher et al. 2002). Einige Konsequenzen des Einsatzes dieses Lichtmodells werden im Folgenden erörtert.

## 2. Schülervorstellungen zum Licht

Wie Untersuchungen an der Universität Potsdam gezeigt haben, kann die Modellfülle beim Modellbildungsprozess von Schülern oder Studenten zu zahlreichen Implikationen führen (Mikelskis et al. 2002). Insbesondere zeigen befragte Schülerinnen und Schüler im Bereich der Optik oft eine Neigung zur Bildung von Hybridmodellen und lassen eine adäquate und notwendige Trennschärfe zwischen den unterschiedlichen Modellen des Lichts vermissen (Horn et al. 2002).

Darüber hinaus präferieren sie Darstellungen auf der Basis des Strahlenmodells mit deutlich determinierten Lichtwegen.

**Schüler 1**: „Ich stelle mir Photonen immer vor wie kleine Teilchen, die eben Quanten sind und die eben sauber durch die Öffnung kommen. Und die vielleicht jetzt eben den oberen Spalt leicht berühren und eben vielleicht in ihrer geraden Flugbahn leicht abgelenkt werden."

**Schüler 2**: „Besonders einfach ist es natürlich, man stellt sich Licht als kleine Teilchen vor.
Da (...) stecken die Teilcheneigenschaften drin, die sich dann in Wellenform weiterbewegen."

***Abb. 1:*** *Äußerungen von Schülern aus Physik-Leistungskursen (13. Jg.) zur Natur des Lichts.*

Diese deterministische Grundhaltung ist in Schüleräußerungen durch die Zuordnung von Teilchenbahnen (siehe Abbildung 1) explizit erkennbar. Der weitaus größte Teil der Schülerinnen und Schüler folgt dieser lebensweltlich begründeten Einstellung.

Die große Attraktivität dieses Konzepts zeigt sich auch durch eine hohe Bereitschaft der Schülerinnen und Schüler, Hybridkonzepte des Lichts zu konstruieren, in denen Bahnen oder Bewegungsrichtungen als wesentlicher Bestandteil enthalten sind. Die Be-cription of sinusoidal trajectories of light particles is a typical example of this process.

schreibung sinusförmiger Trajektorien von Lichtteilchen ist ein typisches Beispiel für diesen Prozess.

### 3. The model of energy flow lines

Models of light which support a deterministic tracking of light paths, particularly accommodate these student conceptions. The model of energy flow lines is one of these student-friendly models.

Energy flow lines technically describe the transport of energy of light from light sources to absorbers. They never cross. Therefore energy flow lines behave in a similar way to field lines of electric or magnetic fields. These characteristic properties of energy flow lines are demonstrated in figure 2 in the example of an arrangement of mirrors.

### 3. Das Modell der Energiestromlinien

Lichtmodelle, die eine deterministische Verfolgbarkeit von Lichtwegen unterstützen, kommen diesen Schülervorstellungen besonders entgegen. Energiestromlinien sind ein solches schülerfreundliches Modell.

Die Energiestromlinien beschreiben formal den Transport von Lichtenergie von Licht-Quellen hin zu Licht-Senken (Absorbern). Sie schneiden sich nicht. Energiestromlinien verhalten sich damit ähnlich wie die Feldlinien elektrischer oder magnetischer Felder. Am Beispiel der Spiegelanordnung in Abbildung 2 kann dieses Verhalten exemplarisch gezeigt werden.

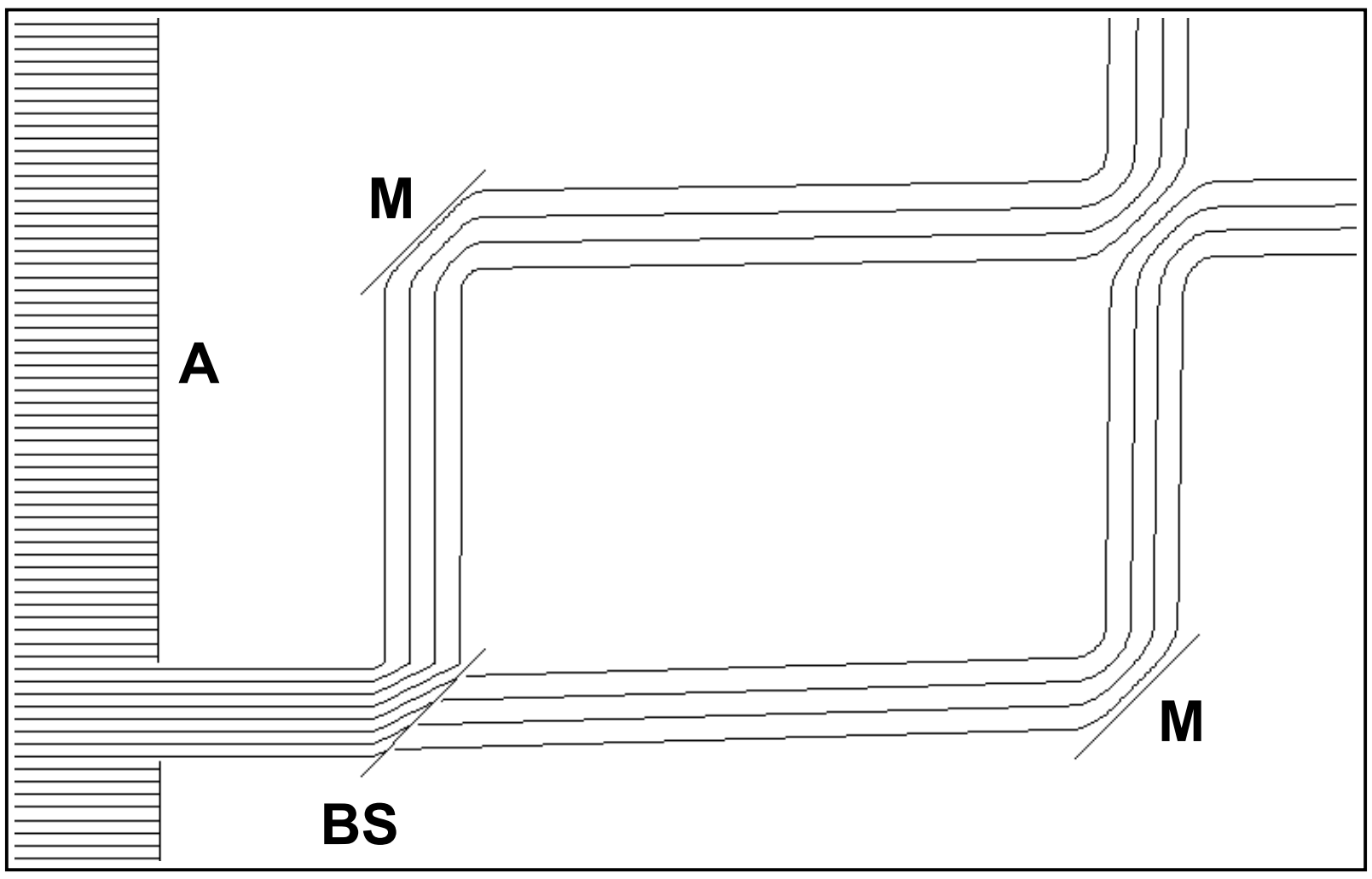

*Parallel light is incident from the left. A fraction of this light will hit a beam splitter and will be reflected at two mirrors in a Mach-Zehnder-like arrangement.*

*Paralleles Licht fällt von links ein. Ein Teil dieses Lichtes trifft auf einen halbdurchlässigen Spiegel und wird in einer Mach-Zehnder-ähnlichen Anordnung an zwei weiteren Spiegeln reflektiert.*

***Fig. 2:*** *Construction of an image of energy flow lines using the open program* Lightlab.
***Abb. 2:*** *Konstruktion eines Energiestromlinienbildes mit Hilfe des Programms* Lightlab. *(Karlsruher Software-Archiv, Software archive Karlsruhe, 2002).*

| | | |
|---|---|---|
| **A** | absorber | Absorberoberfläche |
| **BS** | beam splitter | halbdurchlässiger Spiegel |
| **M** | mirror | Spiegel |

The differences in comparison with the characteristic properties of light rays are clearly visible:

- The trajectories of light run parallel to the surface of the mirror M. Only one trajectory runs right next to the surface of the mirror. The other trajectories of the incident light are subject to a spooky action of distance.
- As the trajectories are not permitted to cross, only the lower half of the pencil of light hitting the slanting beam splitter is able to reach and go through the beam splitter. The upper half of the pencil of light is deflected upwards at great distance from the beam splitter.

Deutlich sind die Unterschiede im Vergleich zum Verhalten von Lichtstrahlen sichtbar:

- Die Trajektorien des Lichts verlaufen parallel zur Spiegeloberfläche M. Nur eine Trajektorie verläuft in unmittelbarer Nähe der Spiegeloberfläche. Die anderen Trajektorien des restlichen Lichts unterliegen einer spukhaften Fernwirkung.
- Da sich die Trajektorien nicht überschneiden dürfen, gelingt nur der unteren Hälfte des Lichtbündels, das auf den schräg platzierten, halbdurchlässigen Spiegel trifft, eine Transmission. Die obere Hälfte des Lichtbündels wird weit vor dem halbdurchlässigen Spiegel nach oben abgelenkt.

- If both pencils of light hit each other again, they will not cross. They will instead be deflected in a region of interaction (see the top right corner of figure 2).

Consequently the light paths of this model construction run in a similar way to the light paths of deterministic quantum mechanics proposed by David Bohm (Bohm & Hiley 1993):

- In the Bohm approach at an infinite high potential barrier the trajectories of quantum objects (like photons or electrons) also largely run parallel to the surface structure of the potential barrier.
- At a finite potential barrier only the light which hits the potential barrier first can penetrate the barrier. The light hitting the potential barrier later is deflected.
- In the region of interaction the Bohm trajectories do not cross (Hiley et al. 2000a, 2000b). They also behave in analogy to figure 2. According to the Bohm interpretation light being detected by a detector at the top of the interaction region unambiguously comes from the left. A detector situated at the right of the interaction region according to Bohm detects only the light which comes from below – provided the light is coherent.

## 4. Didactical consequences

The interpretation of optical effects using the concept of energy flow plays an important part in the area of non imaging optics. Wünscher et al. (2002) discuss numerous practical examples of that. The ***physical content*** of this interpretation therefore is high.

Similarly interesting is the ***epistemological content***. Models and theories in physics are free inventions of the human mind (Poincaré 1905). They can therefore be neither false nor correct. Because of that Wünscher et al. argue that energy flow lines can be interpreted as trajectories and as strictly predetermined paths of light.

The ***internal structure*** of the energy flow line approach closely follows the structure of electrostatics, and at school numerous field line pictures of electrostatic and magnetostatic situations are discussed. Following Wünscher et al. these can be taken as an introduction to the topic of the energy flow of light.

However the ***present-day relevance*** students may assign to a deterministic model of light is didactically problematic. The use of energy flow lines to illustrate optical situations at school can intensify and strengthen the naïve conceptions of students mentioned above. An analysis of these conceptions

- Treffen die beiden Lichtbündel wieder aufeinander, durchdringen sie sich nicht, sondern werden in einer Wechselwirkungszone (rechts oben in Abbildung 2) abgelenkt.

Damit verlaufen die Lichtwege in dieser Modelldarstellung ähnlich wie in der von Bohm vorgeschlagenen deterministischen Quantenmechanik (Bohm & Hiley 1993):

- An einer unendlich hohen Potentialbarriere verlaufen im Bohmschen Modell die Trajektorien von Quantenobjekten wie Photonen oder Elektronen ebenfalls weitgehend parallel zur Oberflächengeometrie der Potentialbarriere.
- An einer endlichen Potentialbarriere tritt nur das zuerst auftreffende Licht durch diese Potentialbarriere hindurch. Das später auftreffendes Licht wird abgelenkt.
- In der Wechselwirkungszone verlaufen die Bohmschen Trajektorien in Analogie zu Abbildung 2 ohne sich zu überschneiden (Hiley et al. 2000a, 2000b). Licht, das ein oben angebrachter Detektor messen würde, kommt in der Interpretation von Bohm eindeutig von links. Ein rechts angebrachter Detektor misst nach Bohm nur das Licht, das von unten kommt – vorausgesetzt, das Licht ist kohärent.

## 4. Didaktische Konsequenzen

Der ***physikalische Gehalt*** einer Interpretation von Lichtfeldern durch Energiestromlinienbilder ist im Bereich der nichtabbildenden Optik nicht zu übersehen. Wünscher und Kollegen (2002) führen diesbezüglich zahlreiche praktische Beispiele an.

Interessant ist auch der ***erkenntnistheoretische Gehalt***. Da alle physikalischen Modelle und Theorien freie Erfindungen des menschlichen Geistes (Poincaré 1905) und deshalb prinzipiell weder wahr noch falsch sind, argumentieren Wünscher und Kollegen, können Energiestromlinien als Trajektorien und damit als fest determinierte Wege, denen das Licht tatsächlich folgt, interpretiert werden.

Die ***innere Struktur*** des Themenkomplexes „Energiestromlinien des Lichts“ knüpft eng an die Elektrostatik und die im Unterricht oft zahlreich erstellten elektro- und magnetostatischen Feldlinienbilder an. Dies kann, wie auch Wünscher und Kollegen hervorheben, als didaktischer Einstieg genutzt werden.

Didaktisch problematisch ist jedoch die oben aufgeführte ***Gegenwartsbedeutung***, die Schülerinnen und Schüler in einem deterministischen Lichtmodell vermeintlich erkennen können. Durch den Einsatz von Energiestrombildern können naive Schülervorstellungen, wie sie oben beschrieben wurden, verstärkt at the University of Potsdam showed that only a minority of students has a meta-conceptual understanding of models. And only this minority of students is able to distinguish precisely between different models and between reality and the world of models.

Without model-oriented lessons aiming to develop and to strengthen the meta-conceptual abilities of students, the presentation of deterministic light models at school can reinforce their naïve conceptions of light. Thus the model of energy flow lines can only be one of many models discussed in physics lessons. All these models should be strictly distinguished from one another.

Indeed, the ***future significance*** of the model of energy flow lines is very questionable. This model can be understood as leading directly to deterministic quantum mechanics. In fact, after discussing the model of energy flow lines at school, the students may comprehend Bohm trajectories easily but these exist only if the light is coherent.

Two non-coherent pencils of light do not interact and cross each other undisturbed (Hiley et al. 2002a). An explanation of this undisturbed penetration of light will then be a considerable didactical problem if later interference is introduced and if students are cognitively fixed upon energy flow lines.

werden. Es zeigte sich bei der Analyse von solchen Vorstellungen an der Universität Potsdam, dass nur eine Minderheit von Schülerinnen und Schülern ein metakonzeptuelles Verständnis besitzt und zwischen unterschiedlichen Modellen sowie zwischen Modell und Realität scharf trennt.

Ohne einen Unterricht, der auf die Ausbildung solcher metakonzeptueller Fähigkeiten zielt, kann die Behandlung von deterministischen Lichtmodellen zur Vertiefung naiver Schülervorstellungen führen. Das Energiestromlinienmodell kann deshalb im Unterricht nur eines von vielen Modellen sein, zwischen denen strikt unterschieden werden sollte.

Sehr zu hinterfragen ist die ***Zukunftsbedeutung***, die dem Energiestromlinienmodell zukommt. Es kann als direkte Vorstufe der Bohmschen Quantenmechanik gedeutet werden. Zwar lassen sich nach Diskussion des Energiestromlinienmodells die Bohmschen Trajektorien viel leichter verstehen, doch diese sind nur bei kohärentem Licht existent.

Bei einer Wechselwirkung zwischen zwei nichtkohärenten Lichtbündeln durchdringen sich die Lichtbündel ungestört (Hiley et al. 2002a). Eine Erklärung dieser ungestörten Durchdringung stellt bei einer späteren Einführung interferenzoptischer Phänomene dann ein erhebliches didaktisches Problem dar, falls die Schüler durch Energiestromlinien geprägt wurden.

## 5. Concluding remarks

Shall we promote a way of teaching physics which aims at a strictly deterministic vision of the world? Or a strictly stochastic one? Bohm quantum mechanics – in my very personal opinion – seems to be mathematically an awkward and structurally an unwieldy theory. Yet how should learners recognize the beautiful and practical if they have not experienced the ugly and unpractical?

The characterisation of energy flow lines as light paths leads didactically straight to the Bohm interpretation of quantum mechanics. Whether through a discussion of energy flow lines of light one can work towards this model in school or avoid or pass is over is ultimately left up to the teacher. For not only are models in physics neither true nor false and a pure invention of the human mind: this is all the more so in the case of didactical models.

## 5. Schlussbetrachtungen

Sollen wir einen Physikunterricht fördern, der auf ein stringent deterministisches Weltbild hinzielt? Oder ein stringent stochastisches? Die Bohmsche Quantenmechanik – und das ist meine sehr subjektive Einschätzung – ist eine mathematisch hässliche und strukturell unhandliche Theorie. Doch wie sollen Lernende das Schöne und Praktische erkennen, wenn sie das Hässliche und Unpraktische nie gesehen haben?

Die Charakterisierung von Energiestromlinien als Lichtwege führt didaktisch zur Bohmschen Quantenmechanik. Ob durch Erörterung von Energiestromlinien beim Licht auf dieses Modell hingearbeitet oder durch ihre Vermeidung dieses Modell übergangen wird, bleibt letztendlich jedem Lehrenden selbst überlassen. Denn nicht nur physikalische Modelle sind weder wahr noch falsch und eine reine Erfindung des menschlichen Geistes – didaktische Modelle sind es erst recht.

## 6. Literaturangaben

**Attachment**

The following energy flow line pictures were presented at the poster session of the GDCP annual conference in Heidelberg. They were constructed with Lightlab (Software archive Karlsruhe 2002).

**Anhang**

Auf der GDCP-Jahrestagung in Heidelberg wurden auch die folgenden mit Lightlab (Karlsruher Software-Archiv 2002) konstruierten Energiestromlinienbilder vorgestellt.

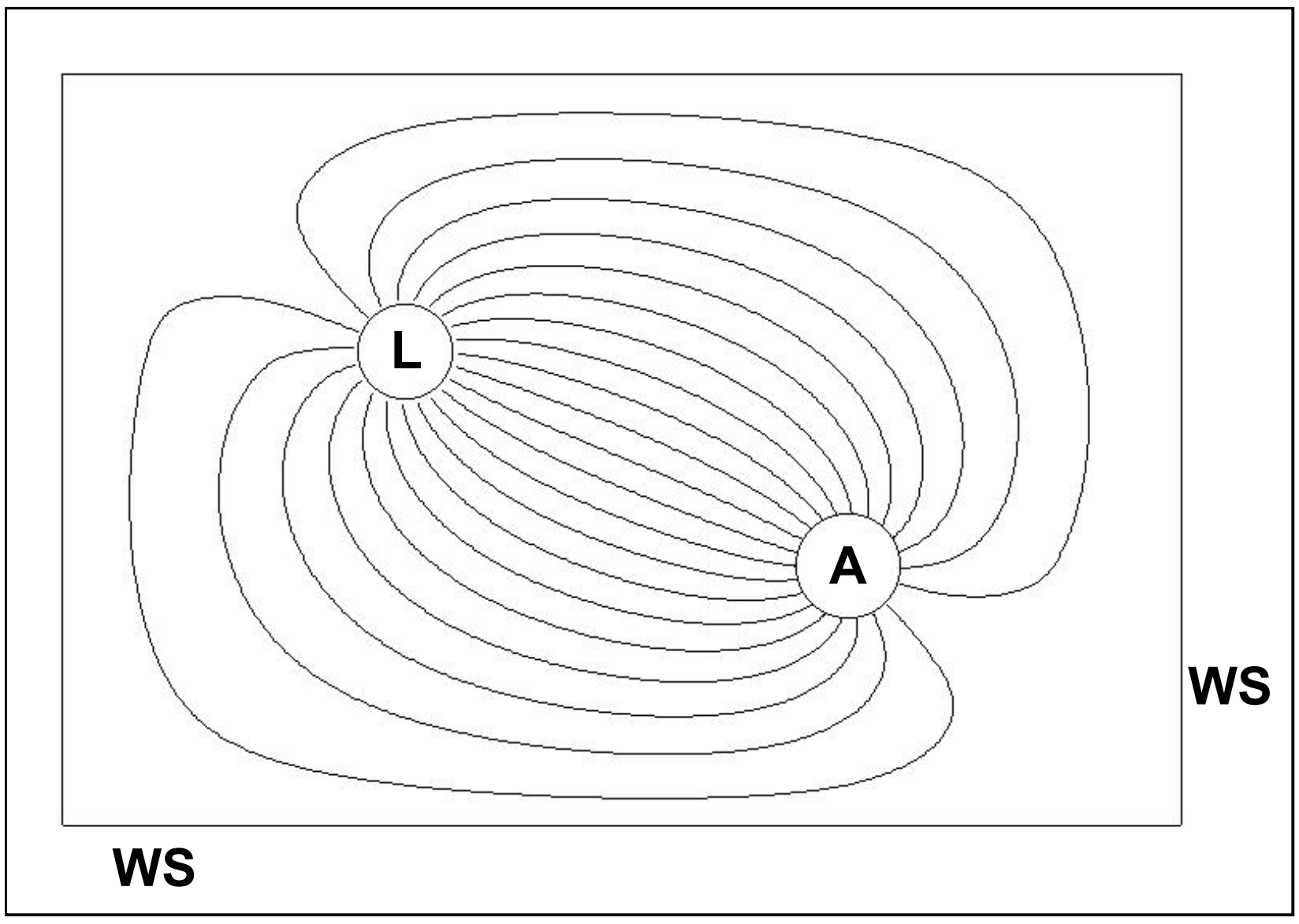

*Figure 3 / Abbildung 3*

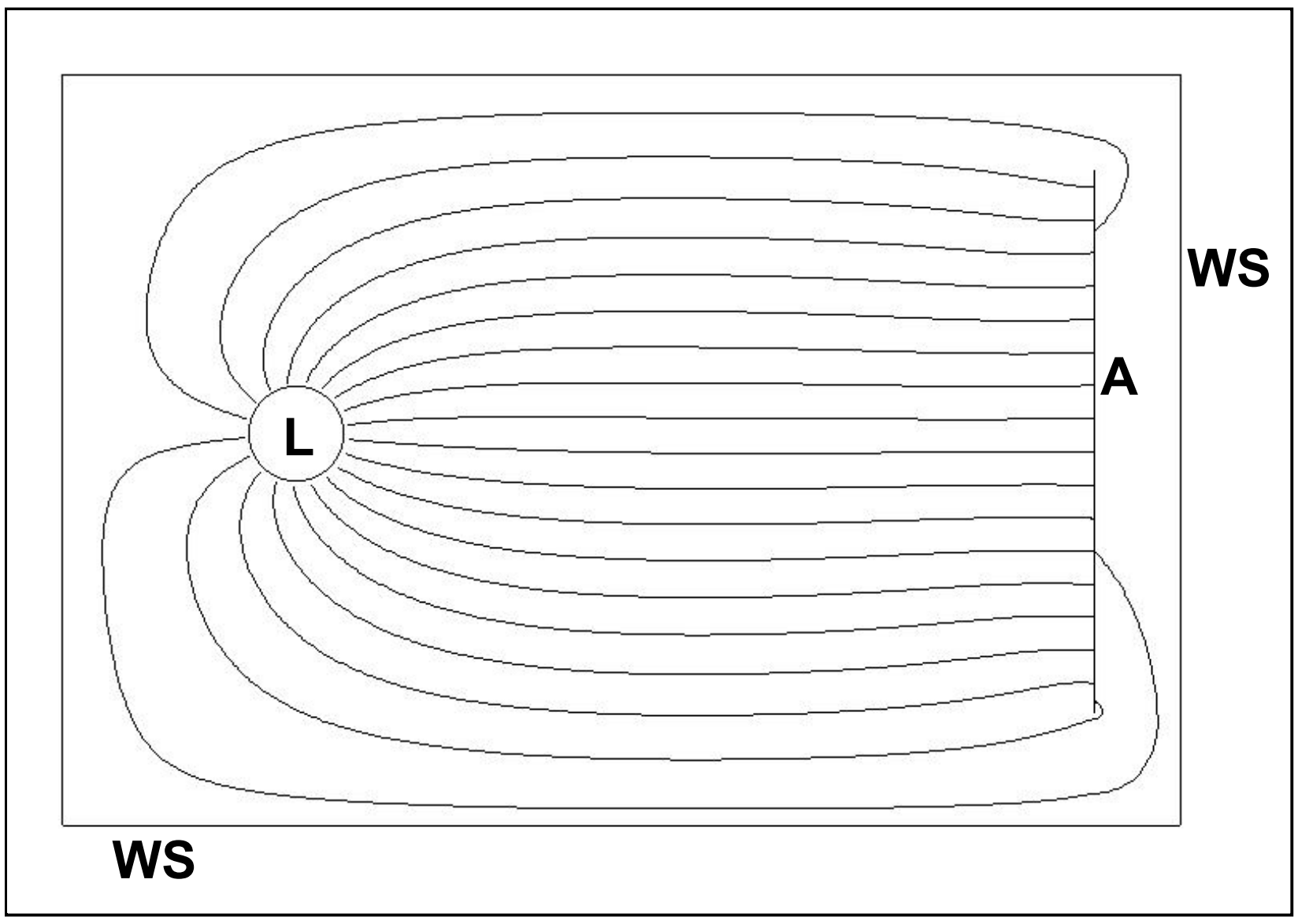

*Figure 4 / Abbildung 4*

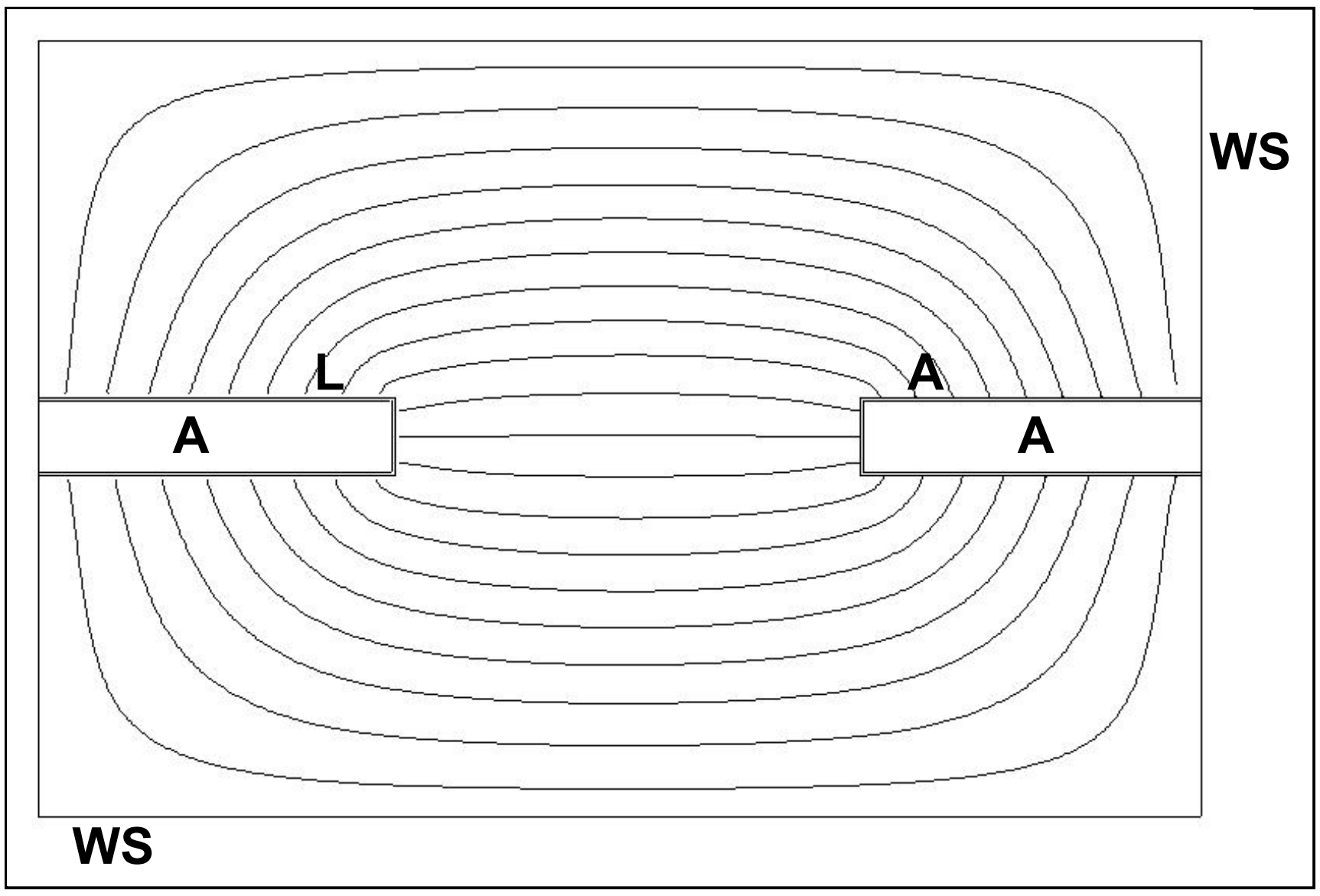

*Figure 5 / Abbildung 5*

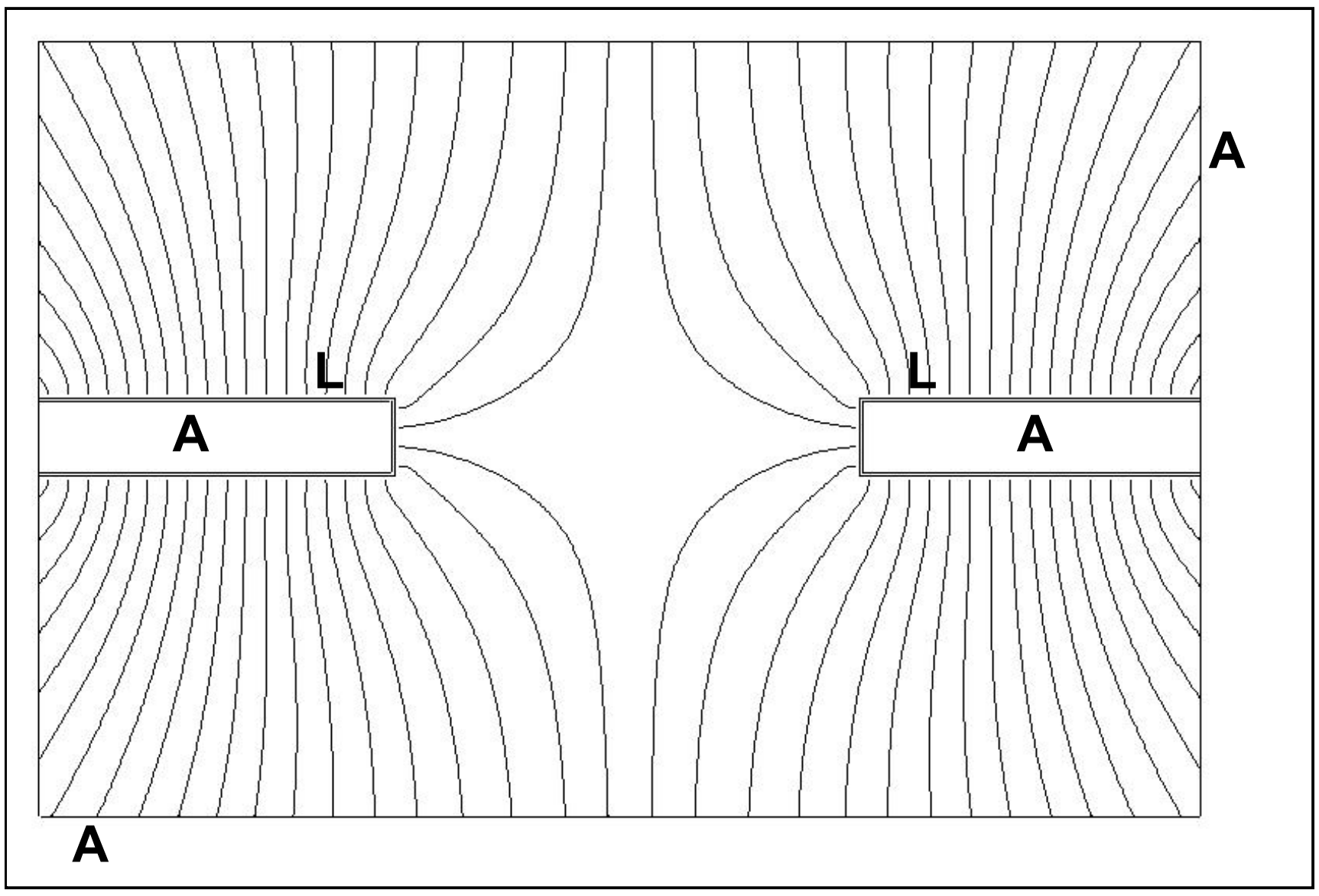

*Figure 6 / Abbildung 6*

***Fig. 3 – Fig. 6 :*** *Energy flow lines which behave in a similar way to field lines of electric or magnetic fields.*
***Abb. 3 – Abb. 6:*** *Energiestromlinien in Anlehnung an Feldlinienbilder der Elektrostatik und der Magnetostatik.*

| | | |
|---|---|---|
| **A** | absorber | Absorberoberfläche |
| **L** | Lambertian light source | Lambertscher Strahler |
| **WS** | white scattering surface (perfect white wall) | diffus streuende Oberfläche (perfekte weiße Wand) |